# Tomonaga–Luttinger liquid in the topological edge channel of multi-layer FeSe

Huimin Zhang[1], Qiang Zou[1], and Lian Li[1]★

[1]Department of Physics and Astronomy, West Virginia University, Morgantown, WV 26506, USA

★Correspondence to: lian.li@mail.wvu.edu

Abstract

A two-dimensional topological insulator exhibits helical edge states topologically protected against single-particle backscattering. Such protection breaks down, however, when electron-electron interactions are significant or when edge reconstruction occurs, leading to suppressed density of states (DOS) at Fermi level that follows universal scaling with temperature and energy, characteristic of Tomonaga–Luttinger liquid (TLL). Here, we grow multilayer FeSe on $SrTiO_3$ by molecular beam epitaxy, and observe robust edge states at both {100}Se and {110}Se steps using scanning tunneling microscopy/spectroscopy. We determine the DOS follows a power-law, resulting in Luttinger parameter $K$ of 0.26 ± 0.02 and 0.43 ± 0.07 for {100}Se and {110}Se edges, respectively. The smaller $K$ for the {100}Se edge also indicates strong correlations, attributed to ferromagnetic ordering likely present due to checkerboard antiferromagnetic fluctuations in FeSe. These results demonstrate TLL in FeSe helical edge channels, providing an exciting model system for novel topological excitations arising from superconductivity and interacting helical edge states.

The defining characteristic of a topological insulator (TI) is the bulk-boundary correspondence that links topological invariants of the bulk band structure to boundary states[1–3]. In two-dimension (2D), the TI is characterized by a single $Z_2$ invariant and helical edge states populated by massless Dirac fermions with spin-momentum locking, which are topologically protected by time-reversal symmetry (TRS) from elastic single-particle backscattering[4–9]. The helical edge states provide robust 1D transport channel with quantized conductance of $2e^2/h$. However, the topological protection breaks down when strong electron-electron interactions are present[10,11], leading to Tomonaga–Luttinger liquid (TLL) with conductance deviating from the ideal value of $2e^2/h$[12]. Recent experiments have provided evidence of conductance plateau at $e^2/h$, or so-called "0.5 anomaly", in the case of HgTe quantum wells[13] when the two helical edges are closely spaced within 100 nm. An alternative explanation for this anomaly is edge reconstruction that can also localize the edge states. Instead of an infinitely sharp edge, reconstruction at edge can spontaneously breaks TRS, giving rise to a conductance that deviates from the perfect quantization[14]. Recent work on Bismuthene epitaxially grown on SiC also shows evidence of TLL in the helical edge channels independent of edge structures[15].

The breaking of TRS, on another hand, can also occur when magnetic ordering or fluctuations are present at the edges. To this end, the superconducting Fe-chalcogenide FeSe has emerged as an ideal platform for probing such phenomena with the interplay of superconductivity, topology, and magnetism[16]. While bulk FeSe exhibits no long-range magnetic order, antiferromagnetic (AFM) spin fluctuations are critical in the understanding of its electronic properties[17]. For single layer FeSe epitaxially grown on SrTiO$_3$(001) (STO), an order-of-magnitude increase in superconducting transition temperature has been observed[18], which likely exhibits unconventional nodeless $d$-wave superconductivity[19–22]. Non-trivial topology has also been theoretically predicted[23,24], and experimentally confirmed by the observation of edge states in single layer FeSe/STO by Scanning Tunneling Microscopy/Spectroscopy (STM/S)[24]. As such, FeSe is an example of AFM TI, where the edges states are protected by a combined symmetry of time-reversal and primitive-lattice-translation[25,26]. For multilayer FeSe, Dirac band dispersion at M point in 50 monolayer (ML) FeSe/STO[27], and topological edge states at nematic domain walls in 20 ML FeSe/STO[26] have also been observed.

In this work, we explore topological edge states in few layer FeSe films grown by Molecular Beam Epitaxy (MBE) (Figure 1a). The tetragonal crystal lattice of FeSe gives rise to two types of step edges: along [100]Se and [$\bar{1}$10]Se directions, as shown in Figure 1b. For both single and multilayer FeSe, the checkerboard (CB) AFM configuration with spin along the $z$ direction yields good agreement with band structures measured by angle-resolved photoemission spectroscopy[24,26]. Under the CB-AFM scenario, the {100}Se edge is ferromagnetically (FM) ordered, while the {110}Se edge is AFM. This difference in magnetic configuration makes it an ideal platform to probe edge-dependent edge states. Using STM/S, we observe robust edge states at both {100}Se and {110}Se step edges. We further determine that their density of states (DOS) follows a power-law as a function of both energy and temperature with exponent $\alpha$ = 1.03 ± 0.10 and 0.37 ± 0.10 for the {100}Se and {110}Se step edges, respectively, confirming the TLL behavior. The corresponding Luttinger parameter $K$ is 0.26 ± 0.02 and 0.43 ± 0.07 for the {100}Se and {110}Se step edges, respectively. The smaller $K$ for the {100}Se edge also indicates strong correlations, attributed to the ferromagnetic ordering likely present on the edge due to the CB-AFM fluctuations in the FeSe system. These observations indicate edge-dependent TLL in the one-dimension helical channels of FeSe films, providing a model system to realize novel topological excitations such as $Z_4$ parafermions arising from coupling of superconductivity and interacting helical edge states[28–30].

**RESULTS AND DISCUSSION**

**Epitaxial growth and electronic structure of FeSe multilayers**. The epitaxial growth of the tetragonal FeSe(001) film follows a mixed mode of step-flow and layer-by-layer under the growth conditions used (See Methods). An STM image showing the typical morphology of FeSe films (2.4 ML) is presented in Figure 1c, where the epitaxial 3rd layer FeSe either follows the step terrace of the STO substrate or forms islands on the 2nd layer. Both yields single layer islands terminated with edges along {100}Se and {110}Se directions. Figure 1d is an atomic resolution image in pseudo-3D mode showing the corner of an FeSe island terminated with a [100]Se edge, and a [$\bar{1}$10]Se edge at 45° angle, as expected from the structure model (Figure 1b). Away from the edge on the upper and lower FeSe layers, a (1x1) structure of the Se lattice is observed. The atomic arrangements at the step edges are also clearly resolved: both the [100]Se and [$\bar{1}$10]Se edges exhibit no reconstructions, though the two rows of the Se lattice nearest to the edge exhibit slightly higher contrast (Figure 1d and Supporting Information Figure S1). At the [100]Se edge, 1× Se lattice is seen, while at the [$\bar{1}$10]Se edge the periodicity is $\sqrt{2}a_{Se}$ consistent with the structural model (Figure 1b).

Note that also apparent on the surface is a network of randomly distributed 1D features with higher contrast, clearly seen on both the 2nd and 3rd layers (Figure 1c and Supporting Information Figure S2a,b), which is attributed to spatially varying tensile strain due to the lattice mismatch between FeSe and the STO substrate[31,32]. The width of the strain modulation is $d \sim 2.2$ nm (Supporting Information Figure S2c). The tensile strain in the FeSe film gradually releases with increasing thickness. Overall, these observations indicate the high quality of the epitaxial FeSe films.

The electronic properties are characterized by dI/dV tunneling spectroscopy, as shown in Figure 1e with two notable features. First, there is a pronounced peak below the Fermi level (marked by black arrows), which appears at around -152 and -146 meV for the 2nd and 3rd layer FeSe. This feature is attributed to the Fe $3d_{z^2}$ bands at $\Gamma$ point[26], which shifts up in energy with increasing tensile strain, as reported in earlier ARPES studies[27]. Second, the DOS is significantly lower around the Fermi energy, giving rise to an apparent "gap" $\Delta = 52$ meV for both layers (Figure 1e and inset).

**Edge-dependent edge states in FeSe multilayers**. Figure 2a shows a topographic image of a 3rd layer FeSe island with alternating {100}Se and {110}Se edges, where dI/dV spectra were taken in the island interior and at the {100}Se and {110}Se edges (Figure 2b). In sharp contrast to the "gapped" spectrum away from the edges, the spectra are V-shaped around the Fermi energy for both types of edges (Figure 2b). To probe the spatial extend of this feature, spatially dependent dI/dV spectra were taken along the arrows marked in Figure 2a from the 3rd to the 2nd layer across the step edges. Away from the [100]Se edge (Figure 2c), the dI/dV spectra show the characteristic Fe $3d_{z^2}$ peak below Fermi level and the "gapped" feature around the Fermi energy for both the 2nd and 3rd layer. However, approaching the [100]Se edge, a V-shape spectrum gradually develops, while the Fe $3d_{z^2}$ peak intensity is gradually suppressed. The depth of the V-shape spectra increases and reaches a maximum right at the edge of the upper layer. Similar results are also observed across the [$\bar{1}$10]Se edge (Figure 2d).

The spatial distribution of these edge states is further revealed by dI/dV conductance maps at select energies. Figure 3a is a topography image where differential conductance maps $g(\mathbf{r}, E)$ were simultaneously taken at the energies indicated (Figure 3b-d). At the step edges, a strong enhancement of DOS is clearly seen at 20 and 40 meV, for all step segments oriented in different directions. Even though the spatial width and intensity varies slightly, the energy window for the emergence of edge states at both {100}Se and {110}Se edges is similar, between 10 and 70 meV, beyond which the edge states disappear (Figure 3d and Supporting Information Figure S3, where more energy dependent dI/dV maps are shown).

To quantify the spatial distribution of the edge states, line profiles across the [100]Se and [$\bar{1}$10]Se edges in the topography image, as well as the $g(\mathbf{r},40\text{meV})$ as marked by the arrows in Figure 3c, are plotted in Figure 3e,f. The edge state distribution can be fitted by a Gaussian function (solid lines). The full width at half maximum, $w$, is found to be 2.14 nm for the [100]Se edge, and 1.88 nm for the [$\bar{1}$10]Se edge. The maximum intensity of the edge states at [110]Se edge is also lower, at 70% of that at [100]Se edge.

The mixed step flow and layer-by-layer growth mode of multilayer FeSe on STO(001) produce a surface morphology consisting of islands of single layer step height, either extending from the existing step edges, or distributed on the terraces (Figure 3g). The majority of the edges are either along the {100}Se or {110}Se directions, connected by either sharp or rounded corners. A dI/dV map $g(\mathbf{r}, 50\text{meV})$ simultaneously obtained is shown in Figure 3h, where pronounced edge states are observed at the edges of all islands and steps, regardless of their shapes. These observations indicate that the edge states are robust across different geometric boundaries, reminiscent of their topological nature as predicted for single layer FeSe[23,24].

**Tomonaga–Luttinger liquid in the edge channels of multilayer FeSe**. The observations presented above firmly establish that there exist robust one-dimensional conducting channels along the edges of multilayer FeSe, where the DOS $\rho(\epsilon)$, as measured by dI/dV tunneling spectroscopy, is suppressed near the Fermi level. To determine the origin of such behavior, we first examine the possibility of a dynamical Coulomb blockade during electron tunneling into nanostructures[33,34]. In this case, the tunneling conductance is determined by junction parameters such as capacitance *C* and resistance *R*, both of which *strongly* depend on the setpoint, i.e. bias voltage *V* or tunneling current *I*. However, this is not observed in experiments when the tunneling current is varied for more than two orders of magnitude from 10 pA to 2 nA (Supporting Information Figure S4). The different setpoints only affected the intensity, while the dI/dV spectra remained V-shaped around the Fermi energy, thus ruling out dynamical Coulomb blockade.

Disorders in 1D metallic systems can also lead to localization and suppression of DOS near the Fermi level[35]. However, edge states are present at well-ordered edges (Figure 3a-b), as well as disordered ones (Supporting Information Figure S5), suggesting disorder is the not the leading factor. In addition, the suppression of DOS should follow an exponential behavior[35], inconsistent with the following analysis that shows the V-shaped dI/dV can be well fit with a power-law[36]:

$$\rho(\epsilon) \propto |eV|^\alpha \quad (1)$$

where *V* is the bias voltage, and $\alpha$ is an edge dependent parameter. Four spectra taken at positions marked in Figure 4a, are shown in Figure 4b, where the bottom spectrum is fitted with the power law. Good agreement is found within an energy window of ± 10 meV with an exponent $\alpha$ = 1.08, except very near the Fermi level. This suggests that electron-electron interactions is responsible for the V-shaped tunneling spectra, which should follow a universal power-law scaling with energy and temperature of the TLL model[37]:

$$\rho(\epsilon, T) \propto T^\alpha \cosh\left(\frac{\epsilon}{2k_BT}\right) \left|\Gamma\left(\frac{\alpha+1}{2} + \frac{i\epsilon}{2\pi k_BT}\right)\right|^2 \quad (2)$$

where $\Gamma$ is the gamma function. At *T* = 4.3 K, all the dI/dV spectra in Figure 4b are well-fitted by Eq. (2) with exponent $\alpha$ ranging from 0.91 to 1.10 within the same energy window. This is further confirmed by temperature dependent studies, as shown in Figure 4c-e. A strong temperature dependence is found, where the zero-bias differential conductance increases as the temperature increases from 4.3 to 27.3 K (Figure 4c).

All spectra in Figure 4c are fitted with Eq. (2) (solid lines), and are then normalized to $1/T^{1.03}$ and plotted as a function of $E/k_BT$, as shown in Figure 4d, where they collapse into a single curve as predicted by TLL theory. The corresponding fitting parameters $\alpha$ at different temperatures are plotted in Figure 4e, showing that they remain nearly constant, at $\alpha$ = 1.03 ± 0.10 for {100}Se step edges. Similar behaviors are observed for {110}Se step edges (Figure 4f-h), albeit with a smaller $\alpha$ = 0.37 ± 0.10. This edge dependence is further confirmed by measurements carried out on other irregular edges, where the value of $\alpha$ exponent is found to vary between the two cases for {100}Se and {110}Se (Supporting Information Figure S6). These observations provide critical evidence for the TLL nature of the 1D channel at {100}Se and {110}Se step edges.

For a helical TLL, the characteristic power-law exponent $\alpha$ is directly related to the Luttinger parameter $K$ by the following equation:

$$\alpha \approx \frac{1}{2}(K + K^{-1} - 2) \qquad (3)$$

With the experimentally determined exponent $\alpha$ = 1.03 ± 0.10 and 0.37 ± 0.10, the helical TLL parameters $K$ are 0.26 ± 0.02 and 0.43 ± 0.07 for the {100}Se and {110}Se edge channels, respectively. This parameter characterizes the strength and sign of electron-electron interactions: $K$ < 1 for repulsion, $K$ > 1 for attraction, and $K$ = 1 for non-interaction. In addition, when $K$ is smaller than 0.25, the system is considered to be under strong interactions, where correlated two-particle backscattering processes are important in the helical edge transport[38]. Hence, the electron-electron interactions are much stronger in the {100}Se edge channel than that in the {110}Se edge channel.

The robust edge states observed here are different from those reported in earlier work for single layer FeSe/STO[24]. In the prior work[24], the single layer FeSe films are in superconducting state, and edge states appear as an enhanced DOS at energies 50-100 meV below the Fermi level. In comparison, the edge states in multilayer FeSe films are the most pronounced at 20-50 meV above the Fermi level (Fig. 3). In addition, the multilayer FeSe films studied here are not superconducting, but in the nematic state with a gap ~ 52 meV near the Fermi level (Figure 1e). Evidence for nematicity is provided in Supporting Information Figure S7, where the four-fold rotational symmetry of FeSe is broken by the formation of stripes consistent with observations in recent work[32,39]. While the origin of such nematicity is still under debate[40], the hybridization of the $d_{xz}$ and $d_{xy}$ bands are generally observed in ARPES measurements[41,42], which can lead to the gap-like feature near the Fermi level (Figure 1e). The gap value of 52 meV remains constant up to the 5$^{th}$ layer FeSe (Supporting Information Figure S8).

The variations of both the spectral shape and TLL parameter $K$ for {100}Se and {110}Se step edges is consistent the CB-AFM spin fluctuations in FeSe. As discussed above, CB-AFM spin fluctuations has been crucial in the understanding of the electronic properties of FeSe, even in the absence of long-range magnetic order[17]. This leads to the {100}Se and {110}Se edges exhibiting different magnetic ordering, as shown in Figure 1b. This different spin configuration likely contributes to different electron-electron interactions, and in turn the different Luttinger parameter $K$. These observations suggest that the multilayer FeSe/STO(001) is an example of AFM TIs, where the edges states are protected by a combined symmetry of time-reversal and primitive-lattice-translation[25,26].

For multilayer films, even-odd step height oscillation of the edge states has been predicted depending on the crystal symmetry[2]. For example, bilayer and trilayer Fe(Se,Te) thin films has been theoretically shown to exhibit different topological behaviors[43]. Such prediction has been experimentally verified in the case of three-dimensional topological crystalline insulator (Pb,Sn)Se, where edge states are only observed at odd surface step edges[44]. In the case of FeSe, while multilayer films can be grown on STO(001), the mixed step flow and layer-by-layer growth only leads to a surface morphology consisting of step edges or islands of single layer height (Figure 3g). Robust edge states are observed at the single layer edges for film thickness up to six layers (Supporting Information Figure 9), the largest thickness studied in this work. Thus the verification of the even-odd step height oscillation of edge states is not possible, even though edge states have been predicted for bilayer height steps, protected by a combination of time-reversal and glide mirror symmetries[26]. In future studies, the growth of FeSe on other substrates such as epitaxial graphene/SiC(0001) will be explored, where multilayer height steps can be obtained[45].

**CONCLUSIONS**

In summary, we have grown multilayer FeSe on STO(001) by MBE and observe robust edge states at both {100}Se and {110}Se step edges. We further identify the DOS follows a universal scaling behavior with both energy and temperature, indicating TLL behavior in the helical edge channels. The smaller Luttinger parameter $K$ for the {100}Se edge indicates strong electron-electron correlations, consistent with the ferromagnetic ordering likely present along the edge due to the CB-AFM fluctuations in the FeSe system. The strong interacting helical edge states coupled with superconductors can support $Z_4$ parafermions[46,47], which are promising for universal, decoherence-free quantum computation. Since FeSe can be easily doped to achieve high temperature superconductivity via either gating[48] or surface electron-doping[49], our findings provide an exciting platform to study the coupling between topological helical edge state and high temperature superconductivity, offering an attractive opportunity to realize these topological excitations.

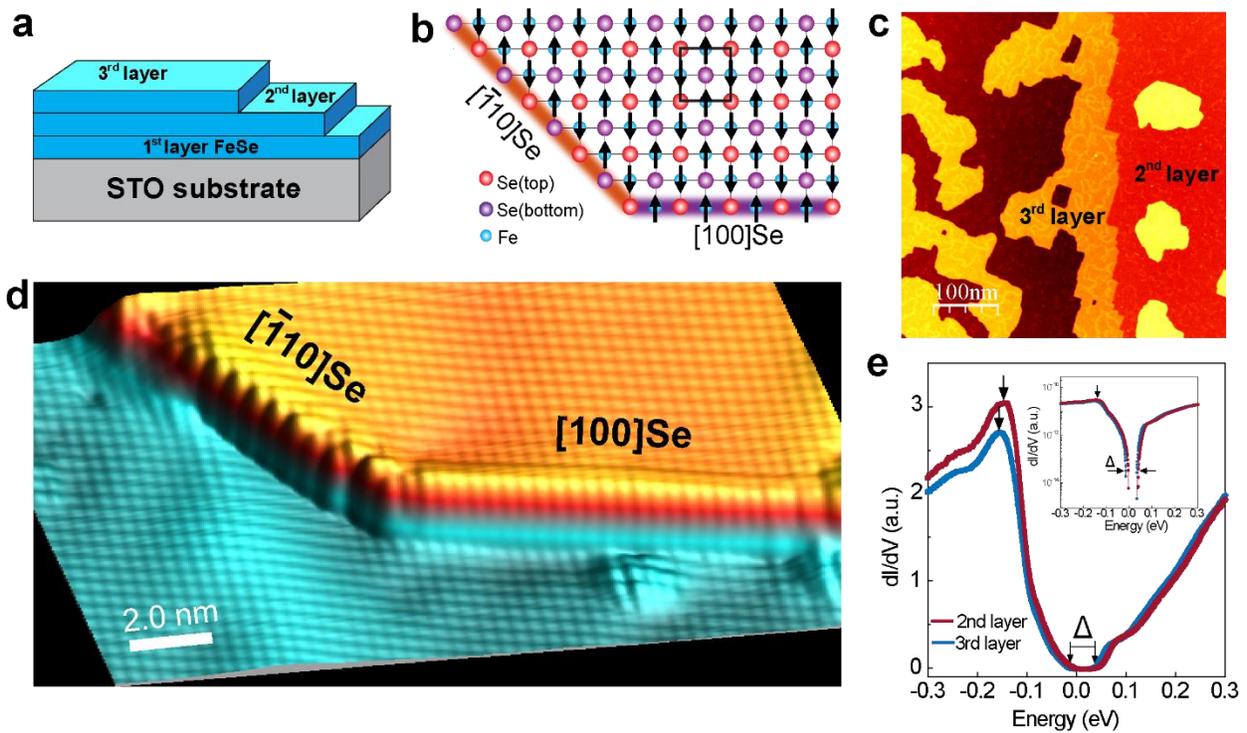

**Figure 1. Topographic and electronic properties of multi-layer FeSe/STO(001) films.** (a) Schematic of multilayer FeSe/STO films. (b) Ball-and-stick model of CB-AFM spin configuration in single layer FeSe. The out-of-plane spin directions are labeled by black arrows. One-unit cell top-layer Se lattice is marked by the black square. The [100]Se step edge (purple) is either perpendicular or parallel to the Se-Se lattice direction, while the [$\bar{1}$10]Se (orange) step edge is at 45º with respect to the Se-Se lattice direction. (c) STM image of a 2.4 ML FeSe film. Setpoint: $V$ = 3.0 V, $I$ = 30 pA. The growth of the 3rd layer FeSe follows either the STO steps or forms islands on the 2nd layer. Setpoint: $V$ = - 3.0 V, $I$ = 30 pA. (d) 3D image of an FeSe island, clearly showing two types of step edges, one along the [100]Se direction and the [$\bar{1}$10]Se direction. Setpoint: $V$ = 100 mV, $I$ = 1.0 nA. (e) Typical dI/dV spectra for the 2nd and the 3rd layer FeSe. The peaks below the Fermi level at - 152 and - 146 meV for the 2nd and 3rd layer, respectively, is attributed to the Fe $3d_z^2$ band. Around the Fermi level, suppressed DOS leads to a "gap" $\Delta$ of 52 meV for both layers, which is more clearly revealed in the logarithm plot (inset).

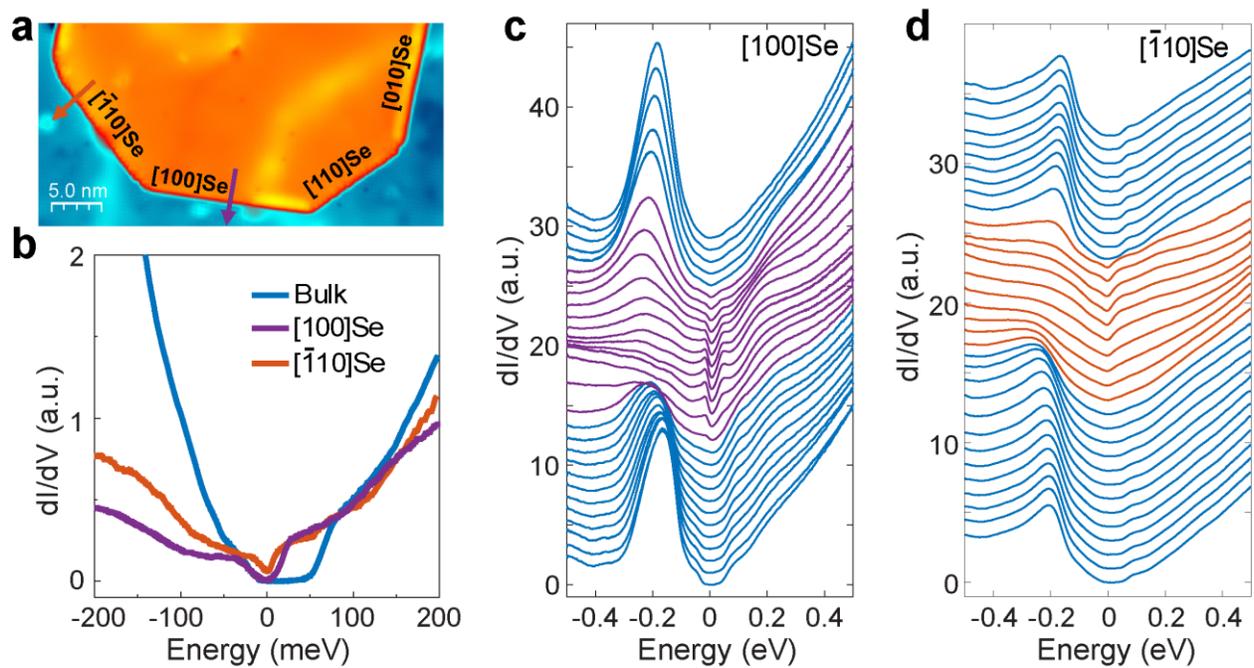

**Figure 2. Spatially resolved dI/dV spectra at the [100]Se and [$\bar{1}$10]Se step edges of a 3ML FeSe film.** (a) STM image of an island on a 3ML FeSe island with alternating step edges as marked. Setpoint: $V$ = 100 mV, $I$ = 1.0 nA. (b) dI/dV spectra taken at [100]Se and [$\bar{1}$10]Se step edges, as well as in the region away from the step edge (bulk). (c) Spatially resolved dI/dV spectra taken across the [100]Se edge along the purple arrow in (a). (d) Spatially resolved dI/dV spectra taken across the [$\bar{1}$10]Se edge along the orange arrow in (a).

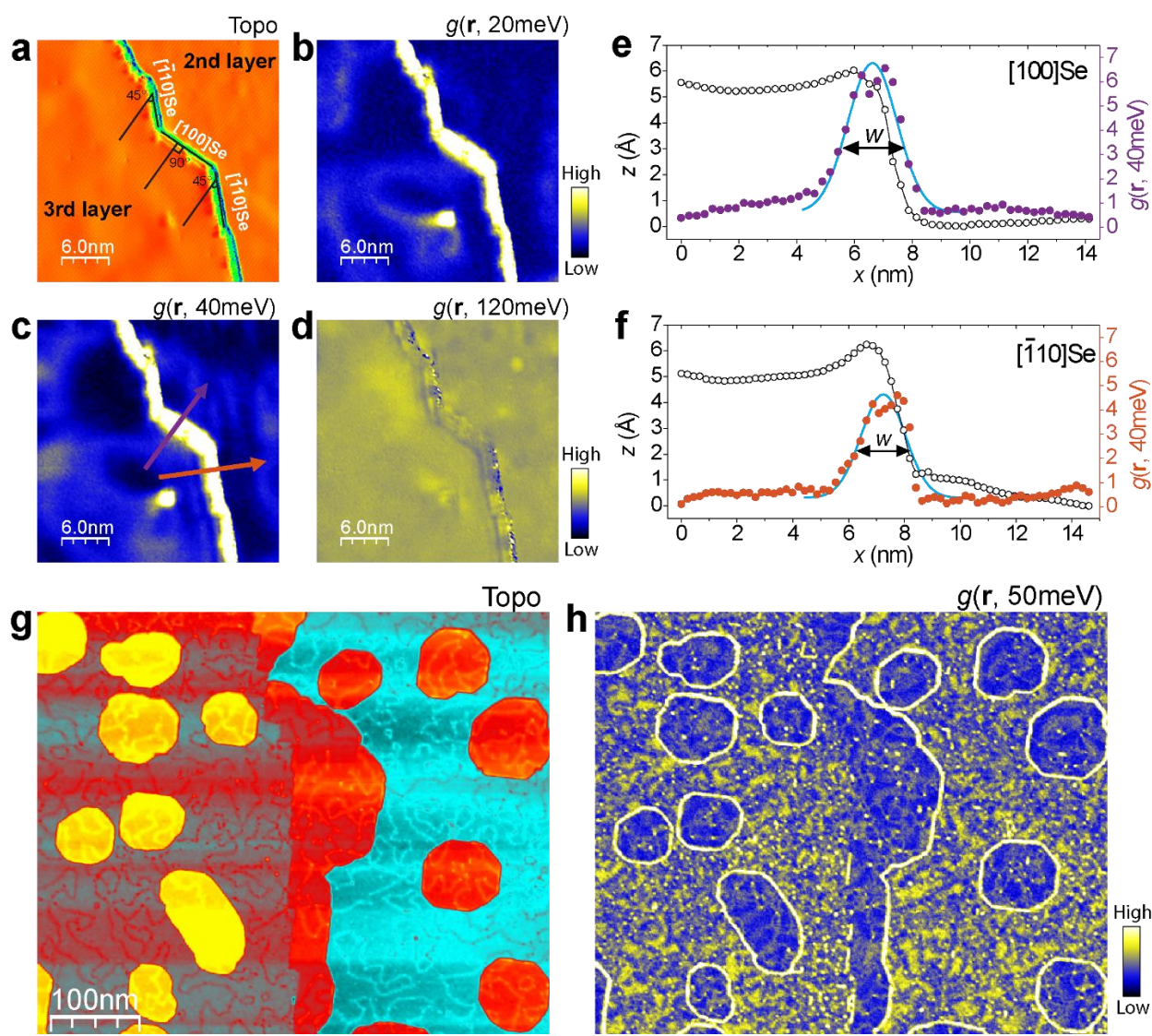

**Figure 3. Spatial distribution of the edge states in FeSe films.** (a) Topography of an FeSe film presented in differential mode, where multiple [100]Se and [$\bar{1}$10]Se step edges are marked. Setpoint: $V$ = 200 mV, $I$ = 0.5 nA. (b)-(d) dI/dV maps, $g(\mathbf{r}, E)$, at various energies indicated. Setpoint: $V$ = 200 mV, $I$ = 0.5 nA, $V_{mod}$ = 5 meV. (e) Line profiles of the topographic and $g(\mathbf{r}, 40\text{meV})$ across the [100]Se step along the purple arrow marked in (c). The spatial distribution of the edge states at the [100]Se edge is determined by fitting the $g(\mathbf{r}, 40\text{meV})$ line profile with a Gaussian function (solid cyan line), giving rise to a full width at half maximum (FWHM), $w$ = 2.14 nm. (f) Line profiles of the topographic and $g(\mathbf{r}, 40\text{meV})$ across the [$\bar{1}$10]Se step along the orange arrow marked in (c). Similar analysis yields a FWHM, $w$ = 1.88 nm. (g) STM image of a 2.4ML FeSe film. Setpoint: $V$

= 200 mV, *I* = 0.1 nA. (h) dI/dV map *g*(**r**, 50 meV) taken simultaneously showing robust edge states along all edges. Setpoint: *V* = 200 mV, *I* = 0.1 nA, $V_{mod}$ = 5 meV.

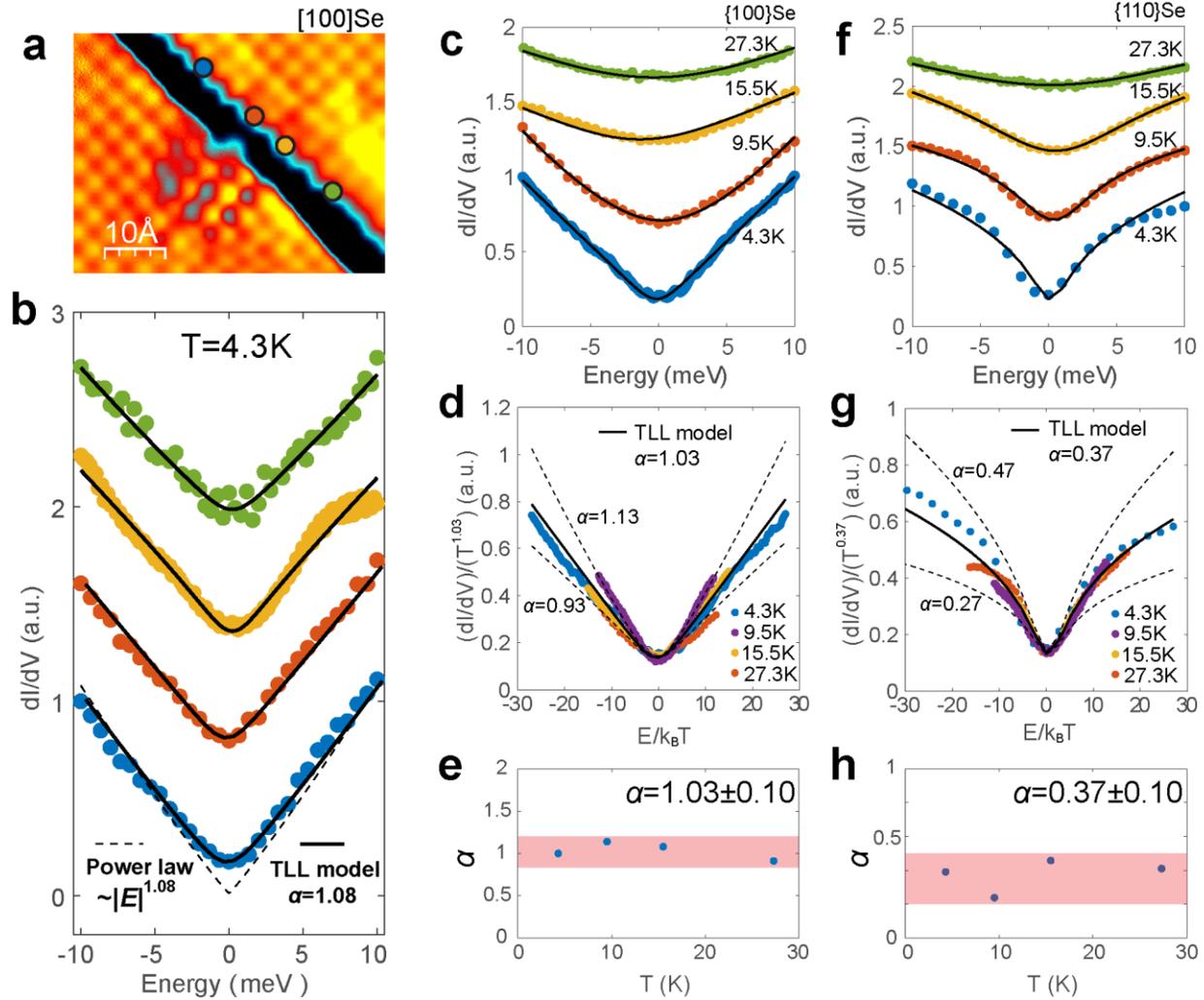

**Figure 4. Tomonaga-Luttinger Liquid behavior in the edge channels of multi-layer FeSe.** (a) STM image of a [100]Se step edge in derivative mode. Setpoint: *V* = 100 mV, *I* = 1.0 nA. (b) dI/dV spectra taken at the positions marked in (a). The black solid lines represent fitting with Eq. (2) with $\alpha$ = 1.08, 1.02, 0.91, and 1.10, from top to bottom, and the dashed line of the bottom spectrum is the power-law fit. (c) Temperature-dependent dI/dV spectra taken at {100}Se edge fitted with Eq. (2) (black solid lines). The spectrum at 4.3, 9.5, 15.5 and 27.3 K is averaged over 4, 15, 17 and 5 spectra, respectively. (d) Universal scaling behavior of the dI/dV spectra as a function of temperature. The dI/dV curves in **c** collapse onto a single universal curve, a hallmark of TLL.

The solid and dashed lines are TLL fittings with the $\alpha$ value indicated. (e) Exponent $\alpha$ as a function of $T$ derived from the TLL fitting in (c), showing an averaged value of $\alpha$ = 1.03 $\pm$ 0.10. (f) Temperature-dependent dI/dV spectra taken at {110}Se edge with TLL fittings (black solid lines). The spectrum at 4.3, 9.5, 15.5 and 27.3 K shown is averaged over 39, 36, 19 and 7 spectra, respectively. (g) Temperature scaling behavior of {110}Se step edge. The solid and dashed lines are TLL fittings with the $\alpha$ value labeled. (h) Exponent $\alpha$ as a function of $T$ derived from the TLL fittings in (f), with an averaged value of $\alpha$ = 0.37 $\pm$ 0.10.

**METHODS**

**Sample preparation.** The multi-layer FeSe films were prepared by MBE on Nb-doped (0.05 wt.%) SrTiO$_3$(001) substrates. To achieve flat surface with step-terrace morphology, the STO substrates were first degassed at 600 °C for 3 hours, followed by annealing at 1000 °C for 1 hour. During the MBE growth, high purity Fe (99.995%), Se (99.9999%) were evaporated from Knudson cells on the STO substrate with a temperature of 300 °C. The steps of FeSe films exhibit two preferable orientations, {100}Se and {110}Se after post growth annealing at 400 °C.

**LT-STM/S characterization.** The STM/S measurements were carried out in a Unisoku ultrahigh vacuum low temperature STM system. A polycrystalline PtIr tip was used and tested on Ag/Si(111) films before the STM/S measurements. dI/dV tunneling spectra were acquired using a standard lock-in technique with a small bias modulation $V_{mod}$ at 732 Hz.

**ASSOCIATED CONTENT**

**Supporting Information**

This material is available free of charge via the internet at http://pubs.acs.org. Atomic resolution STM imaging of [100]Se and [110]Se step edges; STM imaging of strain modulation in multilayer FeSe films; Energy-dependent edge states; Setpoint-dependent dI/dV spectra at [100]Se and [110]Se step edges; Edges states at edges without well-defined orientations; Edge-dependent dI/dV spectra and corresponding TLL fittings; dI/dV maps of nematic boundaries; Spatial and layer-dependent dI/dV spectra in multilayer FeSe; Edge states in FeSe films with different thicknesses (PDF)

**AUTHOR INFORMATION**

**Corresponding Author**

Lian Li − Department of Physics and Astronomy, West Virginia University, Morgantown, West Virginia 26506, United States

Phone: (+1) 304-293-4270; Email: lian.li@mail.wvu.edu

**Authors**

**Huimin Zhang** − Department of Physics and Astronomy, West Virginia University, Morgantown, West Virginia 26506, United States

**Qiang Zou** − Department of Physics and Astronomy, West Virginia University, Morgantown, West Virginia 26506, United States

**Author contributions**

H.Z. and L.L. conceived and organized the study. H.Z. and Z.Q. performed the MBE growth and STM/S measurements. H.Z. and L.L. analyzed the data and wrote the paper. All authors discussed the results and commented on the paper.

**Notes**

The authors declare no competing financial interest.

**ACKNOWLEDGMENTS**

Research supported by the U.S. Department of Energy, Office of Basic Energy Sciences, Division of Materials Sciences and Engineering under Award No. DE-SC0017632.

**TABLE OF CONTENTS**

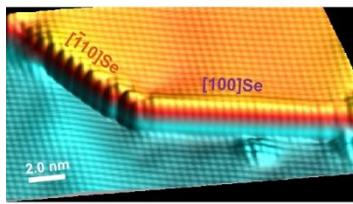 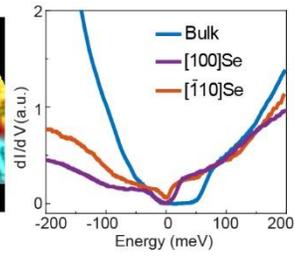

# Supporting Information

## Tomonaga–Luttinger liquid in the topological edge channel of multi-layer FeSe

Huimin Zhang[1], Qiang Zou[1], and Lian Li[1]★

[1]Department of Physics and Astronomy, West Virginia University, Morgantown, WV 26506, USA

★Correspondence to: lian.li@mail.wvu.edu

**Contents:**

Figure S1. Atomic resolution STM imaging of [100]Se and [110]Se step edges

Figure S2. STM imaging of strain modulation in multilayer FeSe films

Figure S3. Energy-dependent edge states

Figure S4. Setpoint-dependent dI/dV spectra at [100]Se and [110]Se step edges

Figure S5. Edges states at edges without well-defined orientations

Figure S6. Edge-dependent dI/dV spectra and corresponding TLL fittings

Figure S7. dI/dV maps of nematic boundaries

Figure S8. Spatial and layer-dependent dI/dV spectra in multilayer FeSe

Figure S9. Edge states in FeSe films with different thicknesses

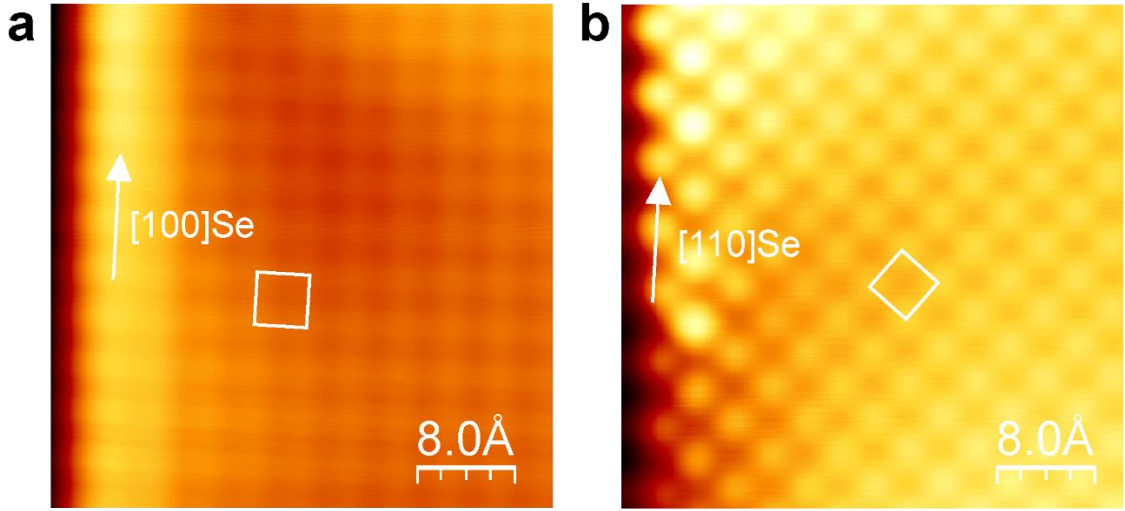

**Figure S1. Atomic structure of step edges of multilayer FeSe films.** (a) Atomic resolution image of the [100]Se step edge. Setpoint: $V$ = 100 mV, $I$ = 1.0 nA. (b) Atomic resolution image of the [110]Se step edge. Setpoint: $V$ = 100 mV, $I$ = 3.0 nA. One-unit cell Se lattice is marked by the white square, and the direction of the step edge is indicated by the white arrow in both images.

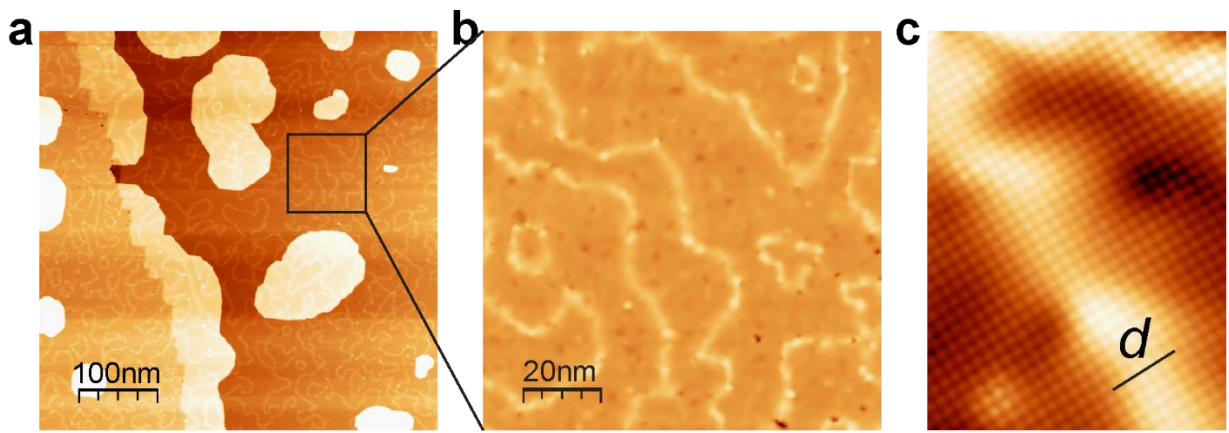

**Figure S2. Strain modulation in multilayer FeSe films.** (a) STM image of a 2.3 ML FeSe film. Setpoint: $V$ = 2.0 V, $I$ = 10 pA. (b) Close-up view of the region outlined in (a). Setpoint: $V$ = 2.0 V, $I$ = 100 pA. (c) Atomic resolution image across a stripe region, where the black line marks the width of the strain modulation $d \sim$ 2.2 nm. Image size: 9 × 12 nm$^2$. Setpoint: $V$ = -1.0 V, $I$ = 9.0 nA.

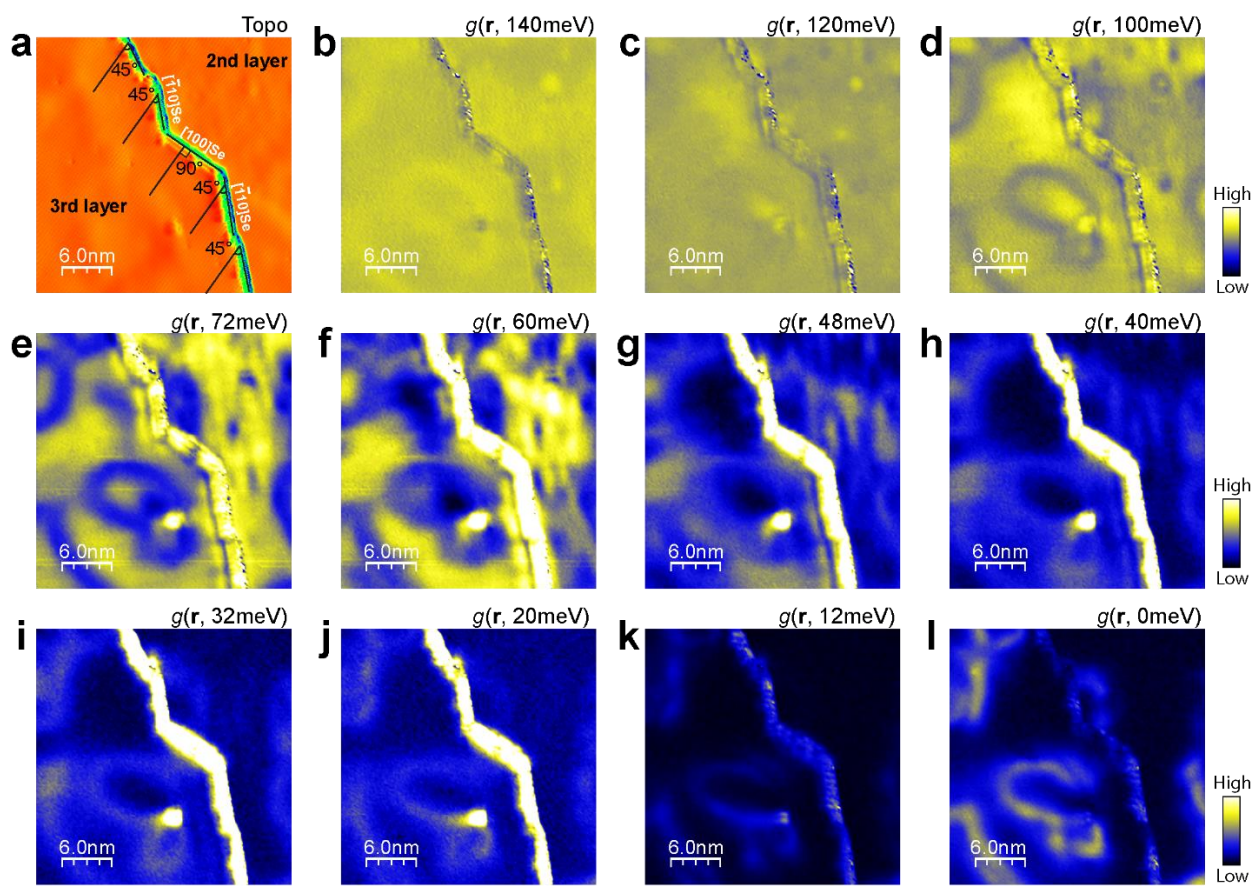

**Figure S3. Energy-dependent edge states at the {100}Se and {110}Se step edges in a 3ML FeSe film.** (a) STM image of a FeSe film presented in differential mode, where multiple {100}Se and {110}Se step edges are marked. Setpoint: $V$ = 200 mV, $I$ = 0.5 nA. (b)-(l) dI/dV maps at energies indicated, $g(\mathbf{r}, E)$. Setpoint: $V$ = 200 mV, $I$ = 0.5 nA, $V_{mod}$ = 5 meV.

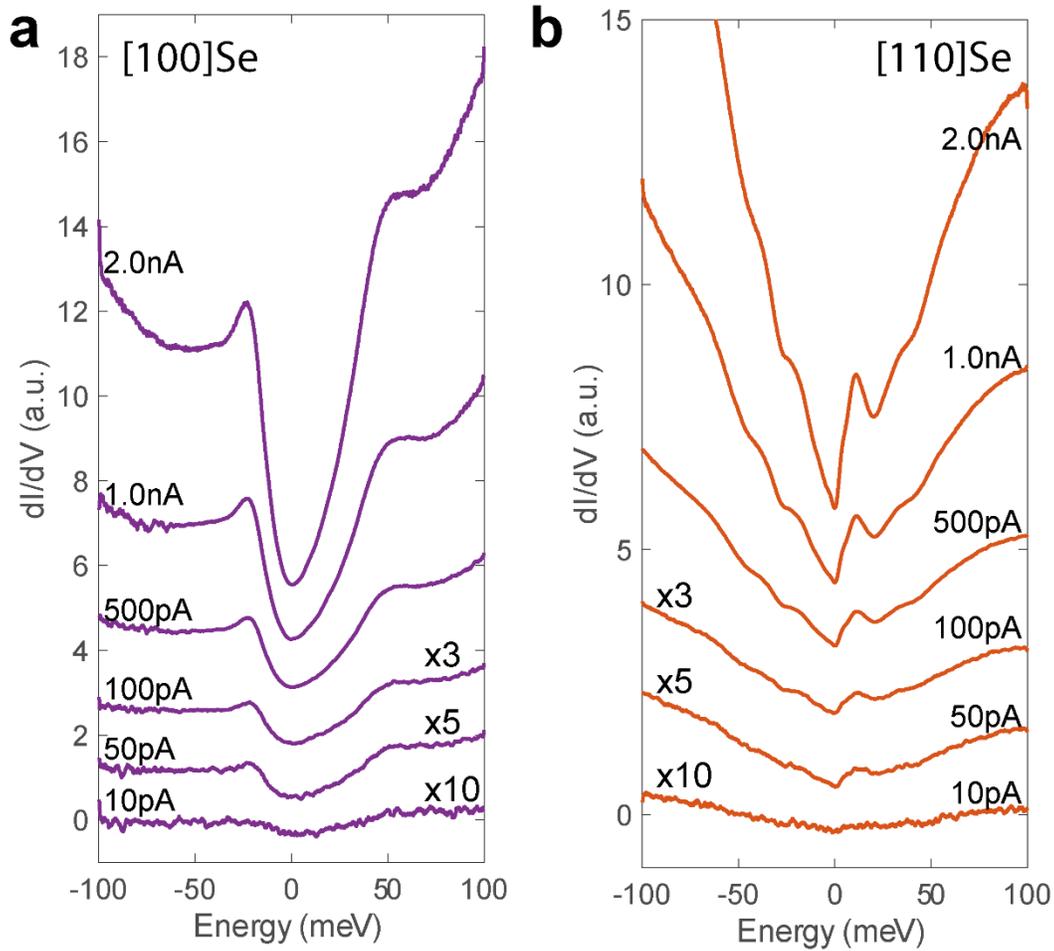

**Figure S4. Setpoint-dependent dI/dV spectra taken at [100]Se and [110]Se step edges.** Setpoint: $V$ = 100 mV, with tunneling current $I$ as indicated. Curves are vertically offset for clarity. The intensity of the dI/dV spectra at 100, 50 and 10 pA are multiplied 3×, 5× and 10×, respectively.

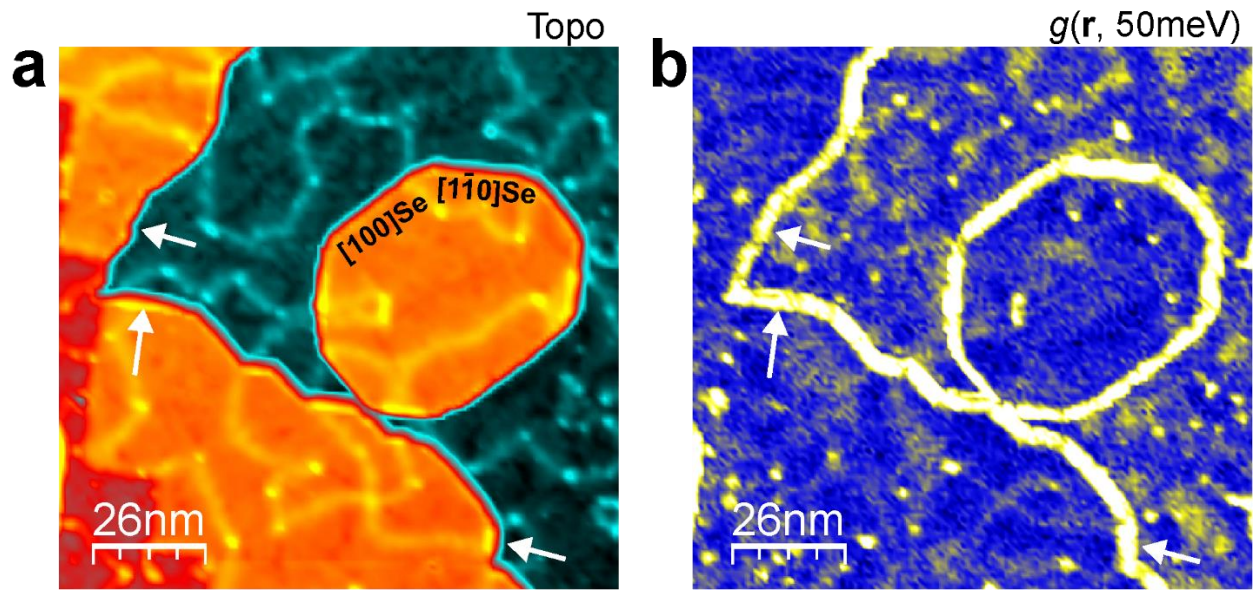

**Figure S5. Edges states at edges without well-defined orientations.** (a) STM image showing ordered and disordered edges of the 3rd layer FeSe. The white arrows mark the disordered edges connecting [100]Se and [1$\bar{1}$0]Se step edges. Setpoint: $V$ = 200 mV, $I$ = 0.1 nA. (b) dI/dV map $g(\mathbf{r}, 50~\text{meV})$ showing robust edge states present along all edges. Setpoint: $V$ = 200 mV, $I$ = 0.1 nA, $V_{mod}$ = 5 meV.

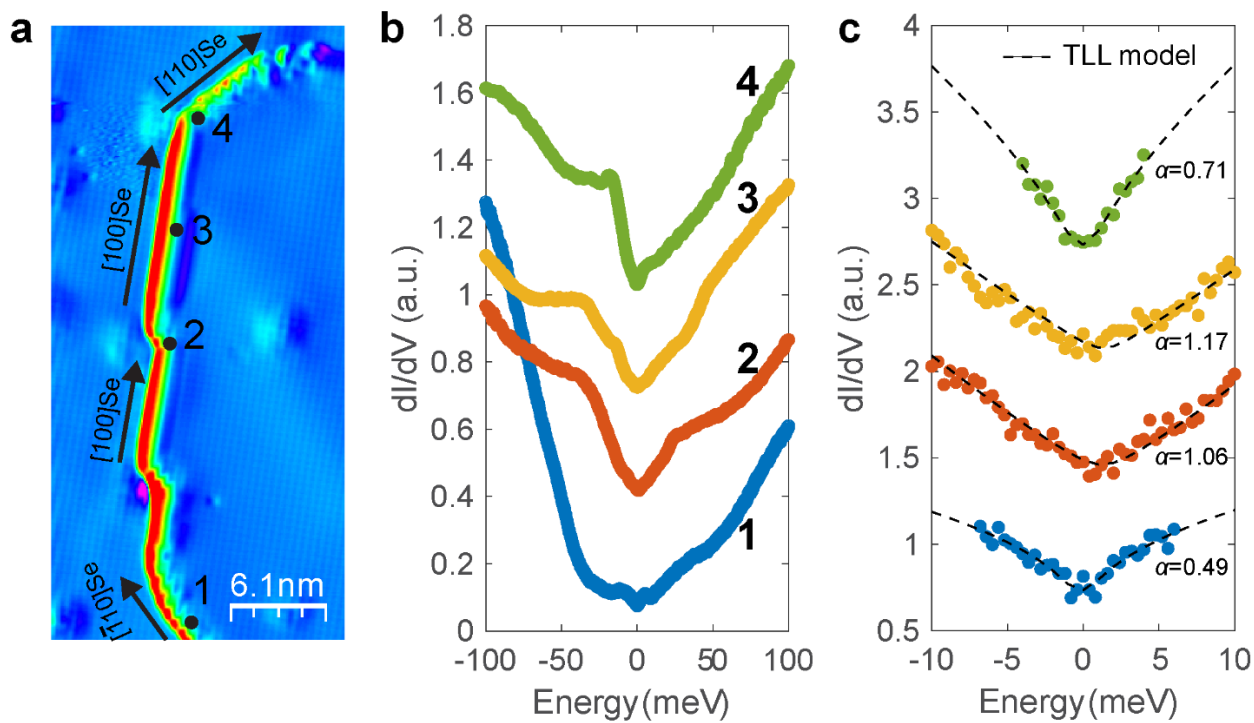

**Figure S6. Edge-dependent dI/dV spectra and corresponding TLL fittings.** (a) STM image of an FeSe island presented in differential mode, where multiple [100]Se and [110]Se step edges are marked. Setpoint: $V$ = 200 mV, $I$ = 0.5 nA. (b) dI/dV spectra taken at locations marked in (a). (c) TLL fitting within the energy window [-10 meV, 10 meV] for point 1-4.

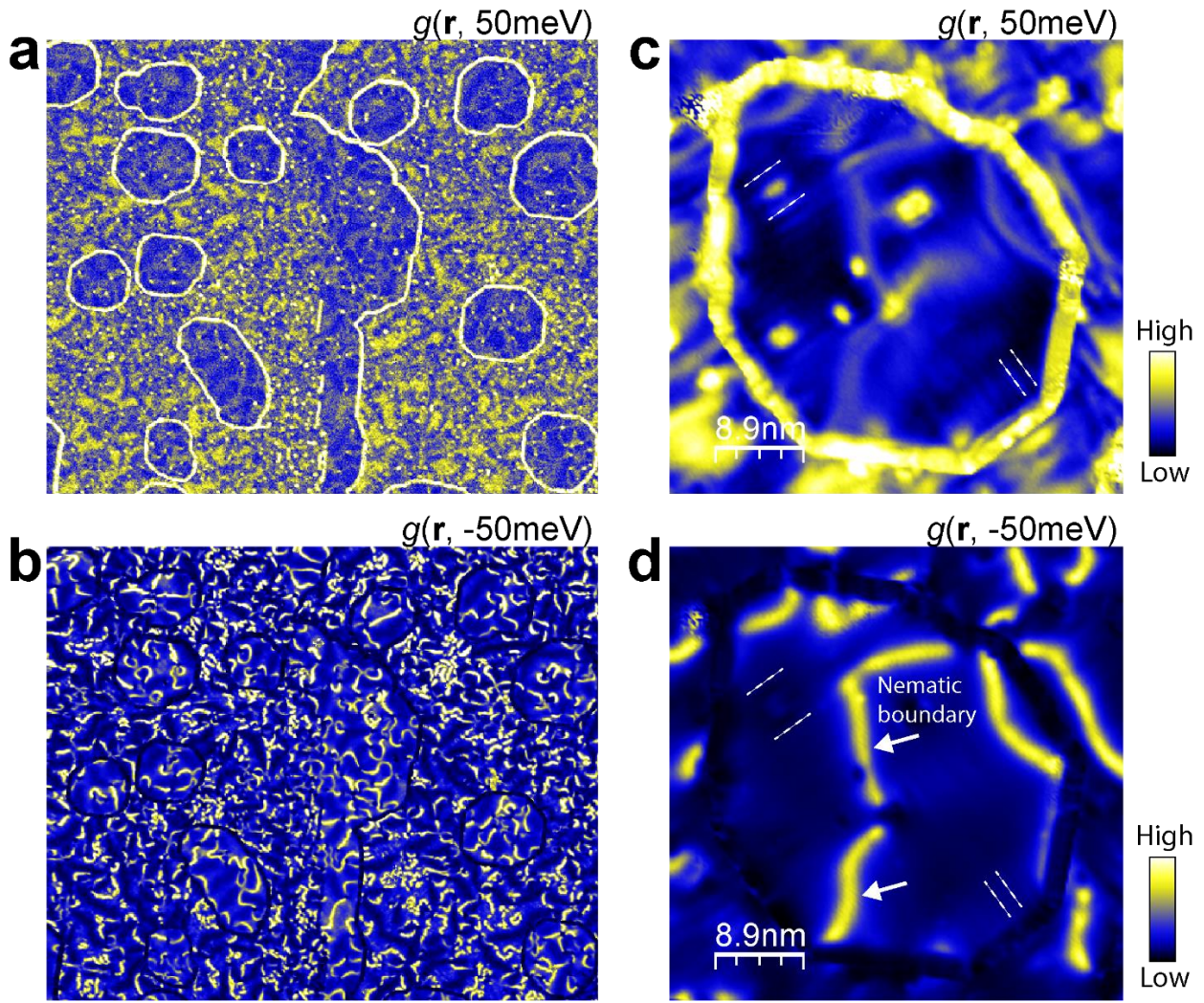

**Figure S7. Evidence for nematicity.** (a)-(b) dI/dV maps $g(\mathbf{r}, 50\text{ meV})$ and $g(\mathbf{r}, -50\text{ meV})$ showing the edge states above the Fermi level and the nematic boundaries below the Fermi level. Setpoint: $V = 200$ mV, $I = 0.1$ nA, $V_{mod} = 5$ meV. The 1D features with higher contrast in (b) are nematic boundaries. (c)-(d) dI/dV maps of a single FeSe island at the energy indicated. Setpoint: $V = 100$ mV, $I = 0.5$ nA, $V_{mod} = 2$ meV. Stripes are observed with their orientations (marked by dashed white lines) rotating 90° across the nematic boundary. The formation of stripes breaks the four-fold rotation symmetry of FeSe, indicating the FeSe films studied here are in the nematic state consistent with recent work[1,2]. The nematic boundary (b) and (d) shows strong contrast at energies below the Fermi level, while the edge states (a) and (c) appear more pronounced above the Fermi level.

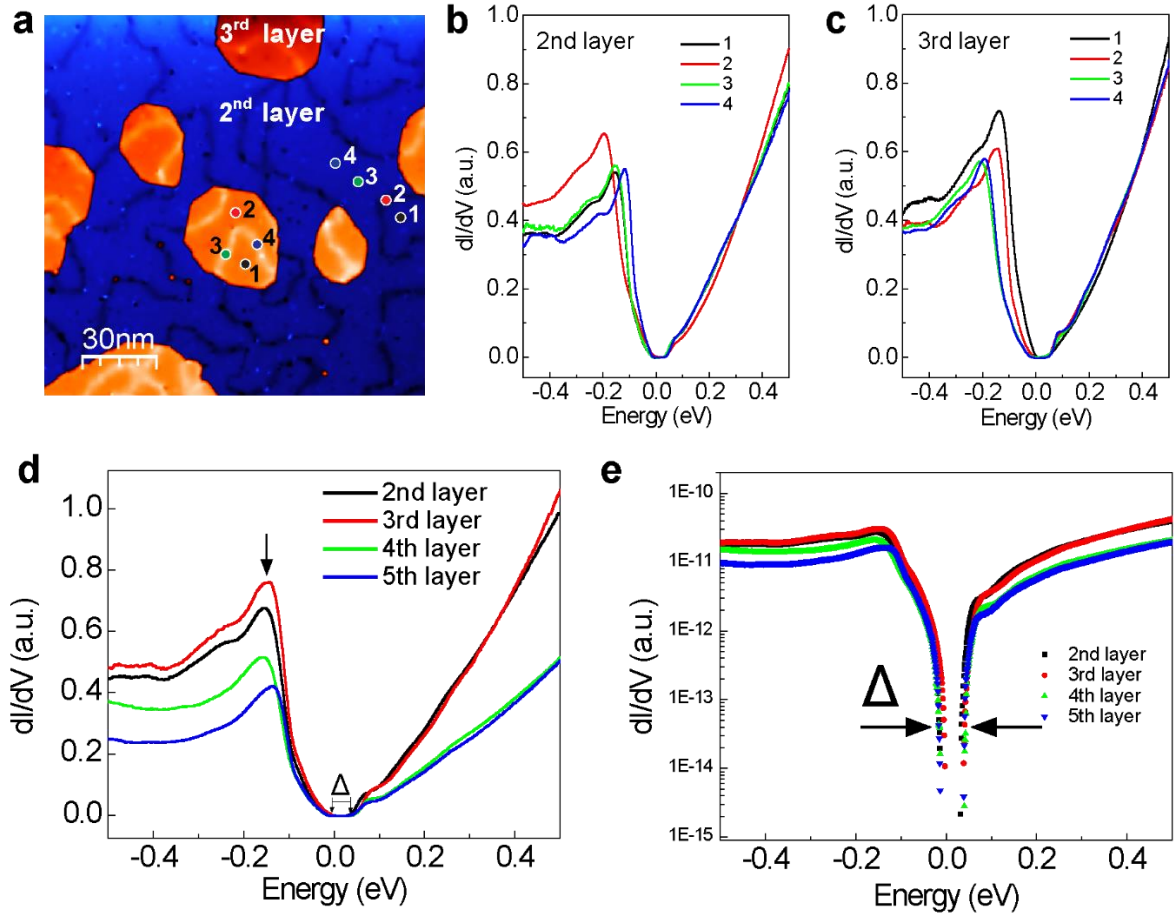

**Figure S8. Spatial and layer-dependent dI/dV spectra in multilayer FeSe.** (a) STM topographic image showing the 3$^{rd}$ layer FeSe islands on the 2$^{nd}$ layer. Setpoint: $V$ = 1.0 V, $I$ = 10 pA. (b) dI/dV spectra taken on the 2$^{nd}$ layer at locations labeled 1-4 in (a). The Fe 3$d_{z^2}$ peak positions are - 152, - 196, - 155 and - 120 meV for spectra 1-4, respectively. (c) dI/dV spectra taken in on a 3$^{rd}$ layer FeSe island with the locations labeled 1-4 in (a). The Fe 3$d_{z^2}$ peak positions are- 135, - 146, - 205 and - 193 meV for spectra 1-4, respectively. These observations indicate the exact position of this peak correlated strongly with the strain boundaries. (d) Typical dI/dV spectra for the FeSe layers between 2-5 ML. Both the pronounced peak below Fermi level and gapped feature around the Fermi level are apparent. The Fe 3$d_z^2$ band is at - 152, - 146, - 157, and - 138 meV for 2$^{nd}$ to 5$^{th}$ layer, respectively. All the spectra are measured away from strain boundaries. (e) Logarithm plot of the dI/dV spectra, where the gap $\Delta$ is 52 meV for all layers.

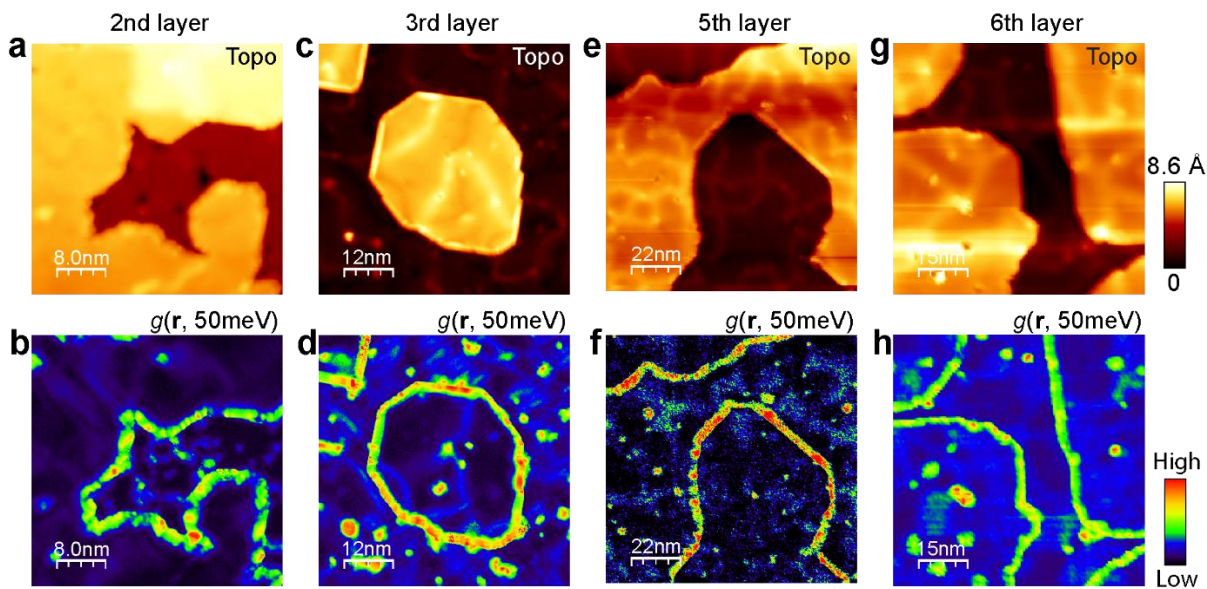

**Figure S9. Edge states in FeSe films with different thickness.** STM topographic image and corresponding dI/dV map $g(\mathbf{r}, 50\text{ meV})$ for a bilayer FeSe film (a)-(b) (Setpoint: $V$ = 200 mV, $I$ = 0.3 nA, $V_{mod}$ = 5 meV); a 3$^{rd}$ layer FeSe film (c)-(d) (Setpoint: $V$ = 100 mV, $I$ = 0.2 nA, $V_{mod}$ = 2 meV); a 5$^{th}$ layer FeSe (e)-(f) (Setpoint: $V$ = 200 mV, $I$ = 0.2 nA, $V_{mod}$ = 1 meV) and a 6$^{th}$ layer FeSe (g)-(h) (Setpoint: $V$ = 200 mV, $I$ = 0.2 nA, $V_{mod}$ = 5 meV).